# Raman vibrational spectra of bulk to monolayer ReS$_2$ with lower symmetry

Yanqing Feng[1]*, Wei Zhou[1]*, Yaojia Wang[1], Jian Zhou[2], Erfu Liu[1], Yajun Fu[1], Zhenhua Ni[3], Xinglong Wu[1], Hongtao Yuan[4,5], Feng Miao[1+], Baigeng Wang[1+], Xiangang Wan[1+] & Dingyu Xing[1]

[1] National Laboratory of Solid State Microstructures, School of Physics, Collaborative Innovation Center of Advanced Microstructures, Nanjing University, Nanjing 210093, China.

[2] Department of Materials Science and Engineering, Nanjing University, Nanjing 210093, China.

[3] Department of Physics, Southeast University, Nanjing 211189, China.

[4] Geballe Laboratory for Advanced Materials, Stanford University, Stanford, California 94305, USA

[5] Stanford Institute for Materials and Energy Sciences, SLAC National Accelerator Laboratory, Menlo Park, California 94025, USA

*these authors contributed equally.

+ Correspondence and requests for materials should be addressed to F. M. (email: miao@nju.edu.cn), B. W. (email: bgwang@nju.edu.cn) or to X. W. (email: xgwan@nju.edu.cn).

ABSTRACT

Lattice structure and symmetry of two-dimensional (2D) layered materials are of key importance to their fundamental mechanical, thermal, electronic and optical properties. Raman spectroscopy, as a convenient and nondestructive tool, however has its limitations on identifying all symmetry allowing Raman modes and determining the corresponding crystal structure of 2D layered materials with high symmetry like graphene and $MoS_2$. Due to lower structural symmetry and extraordinary weak interlayer coupling of $ReS_2$, we successfully identified all 18 first-order Raman active modes for bulk and monolayer $ReS_2$. Without van der Waals (vdW) correction, our local density approximation (LDA) calculations successfully reproduce all the Raman modes. Our calculations also suggest no surface reconstruction effect and the absence of low frequency rigid-layer Raman modes below 100 $cm^{-1}$. Combining with Raman and LDA thus provides a general approach for studying the vibrational and structural properties of 2D layered materials with lower symmetry.

I. INTRODUCTION

2D layered materials have attracted a large amount of research interest due to their rich physics and tremendous application potentials.[1-3] As a nondestructive and powerful technique for characterizing Brillouin zone center (Γ point) phonon properties of materials, Raman spectroscopy yields information about the structure, lattice symmetry, crystal quality and the existence of defects and impurities[4-7]. It has also been widely used to understand the electronic and vibrational properties, as well as their dependence on the thickness of various 2D layered materials.[8-12] While lattice symmetry plays a crucial role in determining their fundamental properties, most experimentally-investigated 2D layered materials have high lattice symmetry, such as graphene and $MoS_2$. For these materials, usually there are low frequency rigid-layer Raman active modes below 100 cm$^{-1}$ in the few-layer and bulk samples due to stacking structure. These modes have very weak intensities, which bring a huge challenge for traditional Raman spectroscopy to experimentally identify all the symmetry allowing Raman active peaks given the strong Rayleigh scattering and special probing approaches needed[12-16]. For example, the crystalline structure of bulk $MoS_2$ belongs to the $D^4_{6h}$ space group and has two S-Mo-S single layers with four Raman active modes, namely $E_{1g}$, $A_{1g}$, $E^1_{2g}$ and $E^2_{2g}$ modes.[13-14] Among these modes, the double-degenerate shear mode ($E^2_{2g}$ mode at 32 cm$^{-1}$) is almost negligible due to strong Rayleigh scattering.[12] In contrast, 2D layered material $ReS_2$ was recently found[17] to have very weak interlayer coupling and quite low symmetry. Using X-ray diffraction, Wildervanck and Jellinek[18] propose that bulk $ReS_2$ crystallizes in a triclinic structure with the space group P$\bar{1}$-$C^1_i$ (No. 2). This structure with one unit cell containing one sandwich had been confirmed by several groups[19-22]. With only one single layer in its unit cell, $ReS_2$ may exhibit novel and intriguing vibrational properties in Raman spectra.

In this article, we systematically studied the Raman spectra of bulk to monolayer $ReS_2$ samples. Due to the one single layer structure in bulk $ReS_2$, $ReS_2$ shows no low frequency (below 100 $cm^{-1}$) rigid-layer vibration Raman modes, sharply contrary to other well-known members of transition metal dichalcogenides (TMDs with the chemical formula $MX_2$, M: transition metal element, X: chalcogenide element). Combining with LDA calculations, we successfully identified all 18 symmetry allowing first-order Raman active peaks for bulk and monolayer $ReS_2$ in experiment. From monolayer to bulk, we observed that only two Raman modes exhibit measurably (~1.0 $cm^{-1}$) thickness induced frequency shifts, while many other modes have only tiny (0.5 or 0.6 $cm^{-1}$) shifts. Such results suggest the very weak interlayer interaction and no surface reconstruction in this layered material. The excellent agreement between the theoretical and experimental Raman results points to such approach as a powerful tool to characterize vibrational and structural properties of 2D layered materials with low symmetry.

## II. METHODS

Raman experiments were performed using a 100x objective in a Horiba-JY T64000 system at ambient conditions. The excitation laser wavelength was 633 nm, with laser power kept as 1 mW and laser spot about 2 µm in diameter. The backscattering geometry was adopted in Raman measurements, with spectrum resolution of 0.6 $cm^{-1}$. We took room temperature spectrum of every sample for 120s, and analyzed Raman peaks of the spectrum by the typical Lorentzian fitting procedure to get the corresponding Raman shift values.

N type single crystals $ReS_2$ were obtained using $Br_2$ assisted chemical vapor transport method (more details in Tongay et al.'s paper[17]). We adopted the typical micromechanical exfoliation method to get few-layer ReS$_2$ flakes from the bulk ReS$_2$ crystal (2D semiconductors) on Si substrate with 300 nm SiO$_2$ by the visual color differences between samples and substrates. Using Zeiss microscope (Scope A1), we could identify the ReS$_2$ flakes, then determine their thickness by an atomic force microscopy (AFM) (Bruker, multimode 8). In FIG.1(a), an optical image of a typical single-layer ReS$_2$ flake was shown. Its thickness was determined to be 0.87 nm by AFM in the ScanAsyst mode (insert of FIG.1(b)). Since the interlayer distance of ReS$_2$ is 0.62 nm, we can thus confirm it is a monolayer flake.

Γ point phonon frequencies calculations for few-layer (labeling as NL, N is the layer number = 1, 2, 3, 4 here) including bulk ReS$_2$ were performed by the small displacement method[23]. Instead of calculating the Raman tensor, based on the group theory, we analyzed the numerical phonon eigenvectors and obtained the Raman active modes. The calculations were performed in the framework of DFT, using the projector augmented wave (PAW)[24] method as implemented in the VASP code (Vienna ab-initio Simulation Package)[25,26]. The basis set cutoff for the wave functions was 550 eV. Different exchange-correlation functions such as the LDA scheme of Perdew and Zunger (CAPZ)[27], the generalized gradient approximation (GGA) schemes of Perdew-Burke (PB), Perdew-Burke-Ernzerhof (PBE)[28] and Perdew-Wang 91 (PW91)[29] were tested. Our numerical results show that among the above mentioned GGA scheme, GGA-PW91 gives the best agreement with the experimental Raman results. Thus we only presented the results of LDA and GGA-PW91. To check the influence of vdW interaction, we also included the vdW-corrected functions (vdW-DFT), Grimme corrections[30] to PBE in the theoretical calculations and listed the results in Table 3 for comparison. The few-layer ReS$_2$ were simulated with a vacuum of 15 Å in the interlayer x direction to ensure negligible interaction between their periodic images. The Brillouin-zone integration was done on uniform Monkhorst-Pack grids of 1

* 24 *24 for few-layer ReS$_2$ and 24 *24 *24 for bulk ReS$_2$. The convergence criterion of self-consistent calculations for ionic relaxations was $10^{-5}$ eV between two consecutive steps. The internal coordinates and lattice constants were fully relaxed for bulk ReS$_2$. For few-layer ReS$_2$, we relaxed the internal coordinates and lattice constants with fixing volume method to avoid the collapse of vacuum. The convergence criterion of relaxation was that the pressures on the lattice unit cell were less than 0.5 kBar and the Hellman-Feynman forces on the ions were less than 0.001 eV/Å.

III. RESULTS AND DISCUSSION

As shown in FIG.1(c), this structure can be thought as a distorted 1T-MX$_2$ dichalcogenides, and the only symmetry operation for this material is the inversion with center located in the middle of Re1-Re3 bond. We started with optimizing the lattice structure. The numerical lattice parameters and independent internal atomic coordinates for bulk ReS$_2$, as well as the experimentally obtained results[20], were listed in Table 1. We find that the standard LDA method successfully reproduces the internal coordinates and lattice constants. While the GGA scheme significantly overestimates $a$ for about 15.1% and the vdW-DFT method gives a smaller value than the LDA and the experimental ones. Each Re of ReS$_2$ has six neighboring S sites, but different with the case in MoS$_2$, the Re-S bond lengths are not equal to each other. The Re1 site bonds with S1, S2, S3, S4, S5 and S7, and the LDA bond lengths are 2.33, 2.36, 2.44, 2.40, 2.34 and 2.42 Å, respectively; Re2-Si (i=1, 2, 3, 4, 6, 8) bond distances are 2.34, 2.46, 2.40, 2.38, 2.31 and 2.51 Å, respectively. But the average lengths are approximately the same for the two independent Re-S bonds (2.38 Å for Re1-S and 2.39 Å for Re2-S), which are shorter than Mo-S bond in MoS$_2$ (2.42 Å[14]). Meanwhile, each Re atom has three neighboring Re sites and there are four Re atoms in the unit cell. They participate in bonding and form a parallelogram-shaped Re4 cluster leading to a single layer structure per unit cell as shown in FIG.1(c).[17] The Re-Re bonding lengths are listed in Table 2 (Similarly, the LDA and vdW-DFT underestimate the Re-Re bonds while the GGA overestimates Re-Re bonds a bit). This forming of Re4 cluster brings significant amount of bond charges between the Re-Re dimers.[17] From FIG.1(c) we also see that different S atoms have completely different environment and span different Re-Re bonds number: S1 spans three metal-metal bonded Re atoms (Re1-Re2, Re1-Re3 and Re2-Re3); S2 and S4 span a single metal-metal bond (Re1-Re4 for S2, Re1-Re2 for S4); and S3 is a little isolated and spans no metal-metal bonds. All these structure properties for bulk $ReS_2$ are quite different from $MX_2$ dichalcogenides such as $1T\text{-}MX_2$[31], $2H\text{-}MX_2$[12,32] and $3R\text{-}MX_2$[33,34].

The non-resonance Raman scattering measurement results on bulk $ReS_2$ are shown in FIG.2(a). Due to the limitation of spectrometer, we can only measure the Raman modes above 100 cm$^{-1}$ shown in FIG.2(a). The strongest peak located at 520.7 cm$^{-1}$ comes from silicon substrate. Similar to the results of Tongay et al.[17], we also found that there are two big Raman peaks located around 150 and 200 cm$^{-1}$. By carefully collecting the signals, we eventually observed 18 modes spreading in 100-450 cm$^{-1}$ range. We marked these observable Raman active modes by the arrows in FIG.2(a), and listed their frequencies in Table 3.

There are 12 atoms in the unit cell, thus bulk $ReS_2$ possesses 36 vibration modes. $ReS_2$ is isomorphic to the point group $C_i$ (the Schoenflies character tables for the point groups can be found in Ref.[35]), and according to decomposition[36] and group theory symmetry analysis, the irreducible representations of the 36 Γ point phonon modes can be written as $\Gamma=18(A_g + A_u)$. There are 18 asymmetric $A_u$ modes and 18 symmetric $A_g$ modes with respect to inversion. Three acoustical modes and infrared active optical modes must be asymmetric under inversion. Thus 15 infrared active and 18 Raman modes can be found in bulk $ReS_2$. It is worth noting that for $ReS_2$ all of the Raman active and infrared active modes are non-degenerate. While for the common $MX_2$ dichalcogenides, such as 2H-$MoS_2$, there exist two-degenerate E symmetry in-plane vibration modes ($E^1_{2g}$ and $E^2_{2g}$)[12,13] besides the non-degenerate $A_g$ modes due to high symmetric lattice structure.

To further answer the questions that whether there are first-order Raman modes located below 100 cm$^{-1}$ and whether 18 observed Raman modes above 100 cm$^{-1}$ are all first-order modes, we performed LDA, vdW-DFT and GGA Γ point phonon calculations based on our optimized structure. It is found that among all the 36 modes, except the three acoustic modes, all the other modes are above 100 cm$^{-1}$ and we presented the numerical frequencies of the 18 $A_g$ modes in Table 3. It is found that both the LDA and vdW-DFT methods calculated frequencies perfectly agree with experimental results. The maximal discrepancy between the LDA and experimental data is only 5.4 cm$^{-1}$ from the $A_g$ mode around 443.4 cm$^{-1}$ and most of the differences are less than 2.0 cm$^{-1}$. Thus we believe that the standard LDA without considering the vdW interaction is already sufficient to describe the geometric structure and the mechanical properties for this weakly interlayer coupled layered material. Tan et al.[10] and Luo et al.[37] also confirmed that the LDA scheme can predict reasonable lattice constants and also reproduce good Raman frequencies matching with the experimental results. On the other hand, the GGA calculations considerably underestimate the Raman frequencies as shown in Table 3. Thus, from here onward, we focus on the LDA results.

Since all the calculated Raman modes frequencies agree with the experimental data very well, we conclude that the 18 experimental peaks, marked in FIG.2(a) and listed in Table 3, are the 18 symmetry allowing first-order Raman modes. And although due to the limit of our equipment, we cannot measure the low frequency Raman mode, we can still conclude that there are no low frequency rigid-layer Raman modes[38] below 100 cm$^{-1}$ due to the fact the unit cell of bulk ReS$_2$ has only one layer. We can further expect that for the MX$_2$ dichalcogenides with distorted 1T structure, such as ReSe$_2$[39] and TcS$_2$[39], such Raman modes should be also absent. On the other hand for those layered compounds whose unit cells have two layers, the low frequency $E_{2g}$ rigid-layer shear Raman vibration modes exist. Some examples include the modes of about 42 cm$^{-1}$ for graphite[10,38], about 52 cm$^{-1}$ in bulk h-BN[40,41], about 32 cm$^{-1}$ in bulk MoS$_2$[12,14] and 22 cm$^{-1}$ in bulk WSe$_2$[12].

By checking the phonon eigenvectors, we now analyze the Raman modes. Although according to the symmetry, all of the Raman active modes of ReS$_2$ belong to A$_g$ mode, we still denote the one with large out-of-plane vibration weights as A$_g$-like mode, with large in-plane vibration weights as E$_g$-like mode and with both strong in-plane and out-of-plane vibration weights as Cp mode[17]. The Re atom is much heavier than S atom, and there are clearly two panels in the phonon spectrum of ReS$_2$ as shown in Ref. 17. The top 24 branches above 250 cm$^{-1}$ are mainly contributed by S motions, while the 12 branches below 250 cm$^{-1}$ basically come from Re vibrations. And our numerical results show that from low frequency to high frequency, the vibration weights of S atoms increase and Re atoms decrease gradually. As shown in Table 3, there are four A$_g$-like modes. Two low frequency A$_g$-like modes (136.8, 144.5 cm$^{-1}$) mainly involve the out-of-plane vibrations of Re atoms, and the high frequency A$_g$-like modes (422.3, 443.4 cm$^{-1}$) mainly involve the out-of-plane vibrations of S atoms. The E$_g$-like modes, located at 153.6, 163.4, 218.2, 238.1 cm$^{-1}$, mainly involve in-plane vibrations of Re atoms; while other two E$_g$-like modes, at 308.5, 312.1 cm$^{-1}$, are mainly in-plane vibrations of S atoms. These frequency distributions of E$_g$-like modes are similar to the results reported by Tongay et al.[17], except one $E_g$-like Raman mode at 218.2 cm$^{-1}$ (213 cm$^{-1}$ in the Raman spectrum of Tongay et al.[17], which was treated as $A_g$-like mode with mostly out-of-plane vibrations). The Cp modes at 275.1 and 282.6 cm$^{-1}$ involve in-plane and out-of-plane vibrations of Re and S atoms; while the Cp modes above 300 cm$^{-1}$ are mainly in-plane and out-of-plane vibrations of S atoms.

We showed several typical vibration modes in FIG.3. The $E_g$-like mode at 153.6 cm$^{-1}$ mainly involves in-plane stretching vibrations of Re-Re bonds along the edge of the Re4 unit, and in-plane vibrations of a pair of S atoms (S1 and S5) atoms as well. The $E_g$-like modes at 163.4, 218.2 cm$^{-1}$ involve in-plane stretching vibrations of Re-Re bonds, and the vibrations of the pair of S atoms which span only one metal-metal bond. The $E_g$-like mode at 238.1 cm$^{-1}$ vibrates mainly along the diagonal line of the Re4 unit with little S atoms vibrations. Different S atoms vibrations span different numbers of the Re-Re dimers leading to different amount of bond charge polarization thus different Raman intensities. Hence, $E_g$-like modes at 153.6, 163.4, 218.2 cm$^{-1}$ peak strong Raman peaks in Raman spectrum shown in FIG.2(a). And the 153.6 cm$^{-1}$ peak has the strongest intensity. $A_g$-like mode at 144.5 cm$^{-1}$ also displays relatively strong intensity as shown in FIG.2(a) due to the participation of the pair of S1 and S5 atoms in-plane stretching vibrations, while $A_g$-like mode at 136.8 cm$^{-1}$ does not involve S atoms vibrations thus shows weak signal.

We also studied the Raman spectra of few and mono-layer ReS$_2$ systems, which have inversion symmetry regardless the layer number is even or odd unlike MoS$_2$[12]. We performed structural optimization and listed the numerical results for monolayer ReS$_2$ in Table 4. To the best of our knowledge, there is no experimental measurement on the lattice parameters for monolayer ReS$_2$, we thus listed previous theoretical data for comparison.[17,42] Our numerical results agree well with the previous theoretical results. As shown in Table 1 and Table 4, the differences between the numerical results of bulk and monolayer ReS$_2$ are quite small.

The Raman spectrum of monolayer ReS$_2$ was shown in FIG.2(b) and the observable 18 Raman active peaks were marked in arrows. We listed the concrete frequencies in Table 3. The monolayer ReS$_2$ shows very similar Raman spectrum to that of bulk one in the 100-450 cm$^{-1}$ range without new peaks indicating that the symmetry is remained from bulk to monolayer. Our experimental results showed that the first A$_g$-like mode exhibits an obvious red shift as the thickness decreases from bulk to monolayer and softens by 1.1 cm$^{-1}$. The sixteenth Cp mode exhibits about 1.0 cm$^{-1}$ blue shift. Many other modes have tiny (0.5 or 0.6 cm$^{-1}$) thickness induced frequency shifts. And our theoretical calculations shown in Table 3 gave the same frequency shifts. These results indicate the absence of surface reconstruction. In contrast, in MoS$_2$ systems, the surface reconstruction[43] softens the A$_g$ mode of the topmost layer by 25 cm$^{-1}$.[8]

We next conducted a systematic Raman spectra study of few-layer (from 2L to 4L) ReS$_2$ flakes, which were also shown in FIG.2(b). It shows similar Raman spectra to that of bulk and monolayer ReS$_2$, with 18 corresponding observable Raman active modes in the 100-450 cm$^{-1}$ range. As the layer number N increases, there are 12N atoms and 36N vibration modes in NL, with the point group $C_i$, Γ=18N(A$_g$ + A$_u$), and there should be 18N non-degenerate first-order Raman A$_g$ modes. But the differences of the increased Raman modes between NL (N ≥ 2) and monolayer are tiny beyond the spectrum resolution, together with the tiny thickness induced frequency shifts and structural optimization and Raman frequencies differences above discussions, suggesting the ultra-weak interlayer interaction. Our experiment and numerical results are consistent with recent work[17,44], where the monolayer behavior and ultra-weak interlayer interaction in bulk ReS$_2$ due to electronic and vibrational decoupling were explored.

Meanwhile, from bulk to monolayer, we do not notice any regular and distinct thickness dependent tendency of the line widths and the peak intensities of all the 18 Raman active peaks.

It is worth mentioning that there is also work reporting another different crystal structure[45]. Very recently the low frequency interlayer modes had been reported in ReSe$_2$[46], but it had been suggested that ReS$_2$ is not isostructural with ReSe$_2$ (See Ref. 20). Based on the structure with 24 atoms per unit cell[45], we thus also performed the calculations. The most distinct difference between the Raman spectra from one sandwich per unit cell[18-22] and two sandwiches per unit cell[45] is the presence of a low frequency interlayer mode (around 35.8 cm$^{-1}$), which is much below the spectrometer low limit of 100 cm$^{-1}$ of our commercial Horiba-JY T64000 system. Thus experimentally searching the low frequency Raman mode will finally unambiguously identify the structure of ReS$_2$ in the future.

## IV. CONCLUSION

In conclusion, we presented both DFT and experimental Raman scattering studies on lattice vibrations of bulk to monolayer ReS$_2$, which is a 2D TMD material with low lattice symmetry. Combining with LDA calculations, we successfully identified all symmetry allowing Raman modes for both bulk and monolayer ReS$_2$. We find that the low frequency rigid-layer vibration modes are nonexistent in this low symmetry bulk ReS$_2$ and due to the rather weak interlayer interaction, the thickness induced frequency shifts are rather small and there is no surface reconstruction. We believe combining with Raman measurement and LDA calculations is an efficient way for studying the vibrational and structural properties of 2D layered materials with low symmetry.

**Figure 1.** (Color online) (a) A typical microscopic image of monolayer (in the red circle) and multilayer ReS$_2$ flakes on Si substrate with 300 nm thick SiO$_2$. (b) An AFM image of the monolayer ReS$_2$ flake in the red circle of (a). Insert: the height probe profile of the monolayer ReS$_2$, showing the monolayer ReS$_2$ height of 0.742 nm. (c) The lattice structure of bulk ReS$_2$. Left figure, primitive cell of bulk ReS$_2$ perspective drawing down the interlayer lattice vector a. The vector b and c are in-plane lattice vectors. Right figure, periodical lattice structure of bulk ReS$_2$. The parallelepiped is the primitive cell of bulk ReS$_2$.

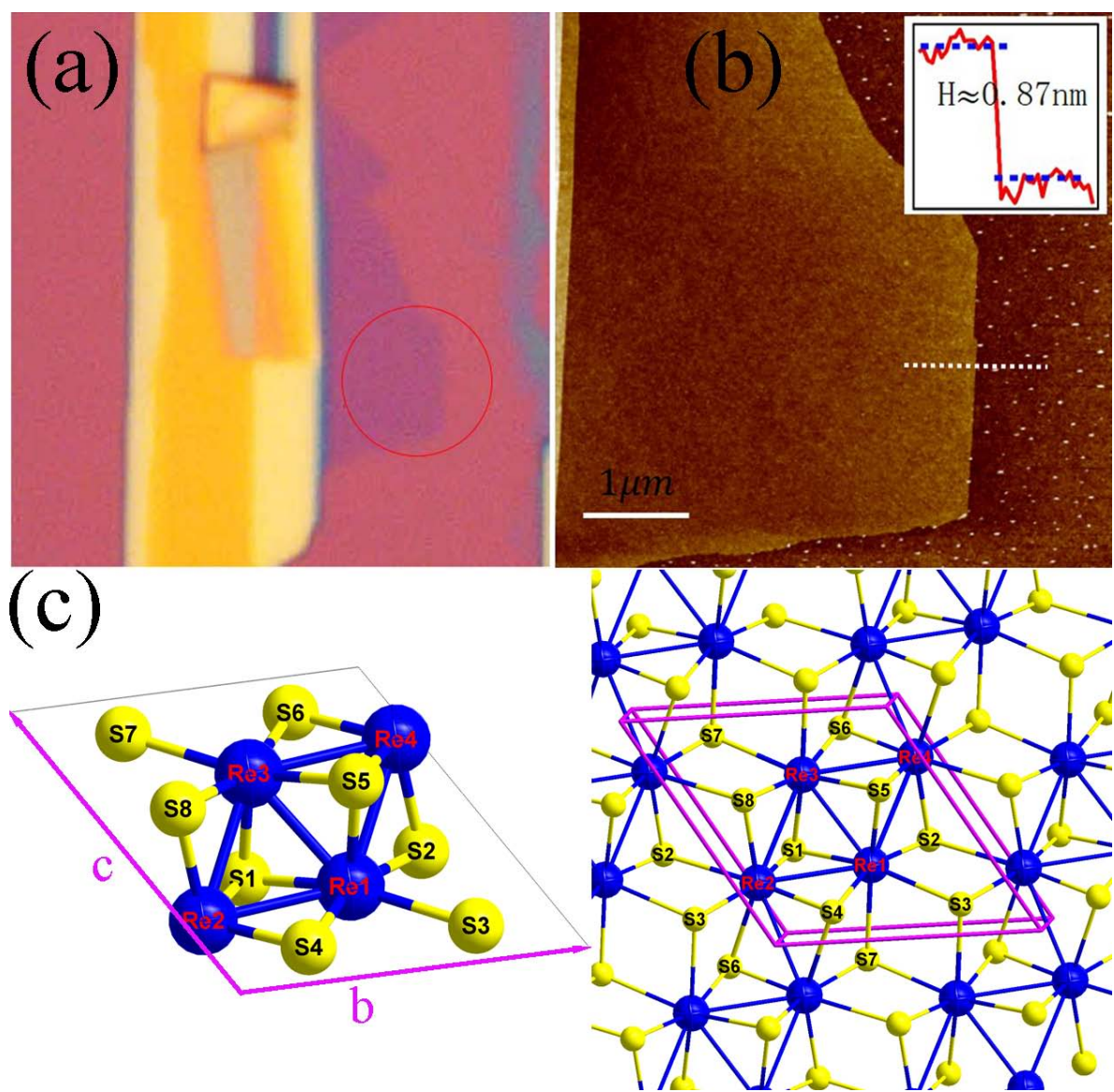

**Figure 2.** (a) Raman spectrum of bulk ReS$_2$. The arrows mark the concrete positions of 18 Raman active modes of bulk ReS$_2$. (b) Raman spectra of few-layer ReS$_2$. The arrows mark the concrete positions of 18 Raman active modes of monolayer ReS$_2$.

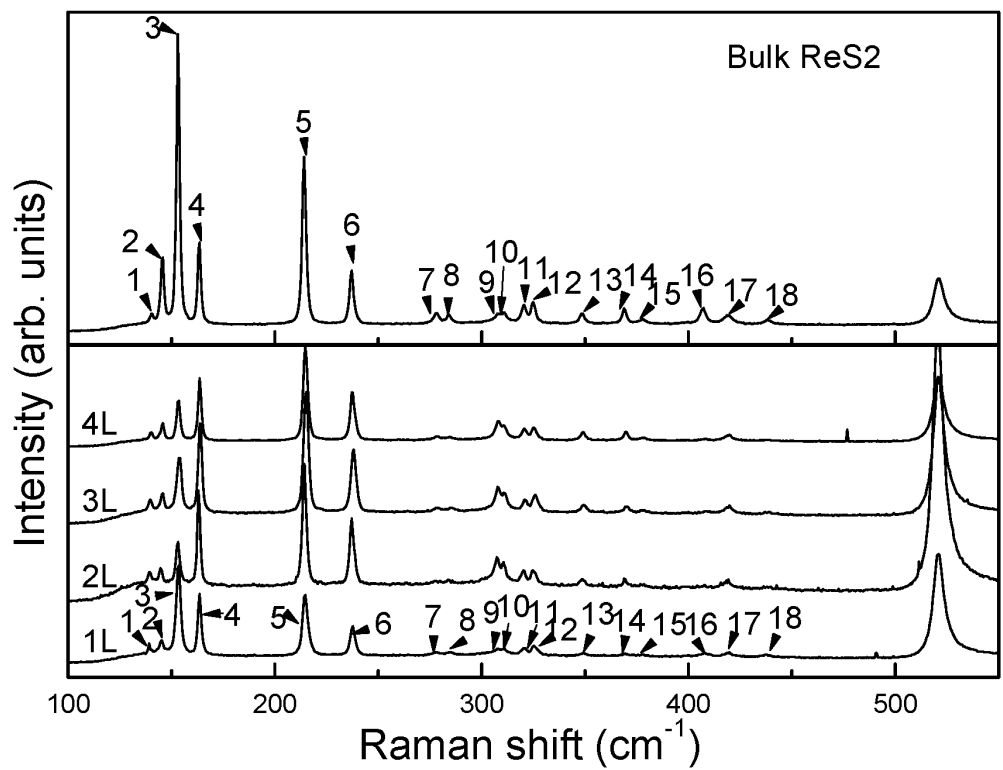

**Figure 3.** (Color online) Four $E_g$-like vibration modes at 153.6, 163.4, 218.2, 238.1 cm$^{-1}$ and two $A_g$-like vibration modes at 136.8 and 144.5 cm$^{-1}$ for bulk ReS$_2$, from LDA theoretical calculations and analysis of the vibration eigenvectors. The lengths of the arrows are proportional to the modular of the phonon eigenvectors with the length weights less than 15 percentage ignored.

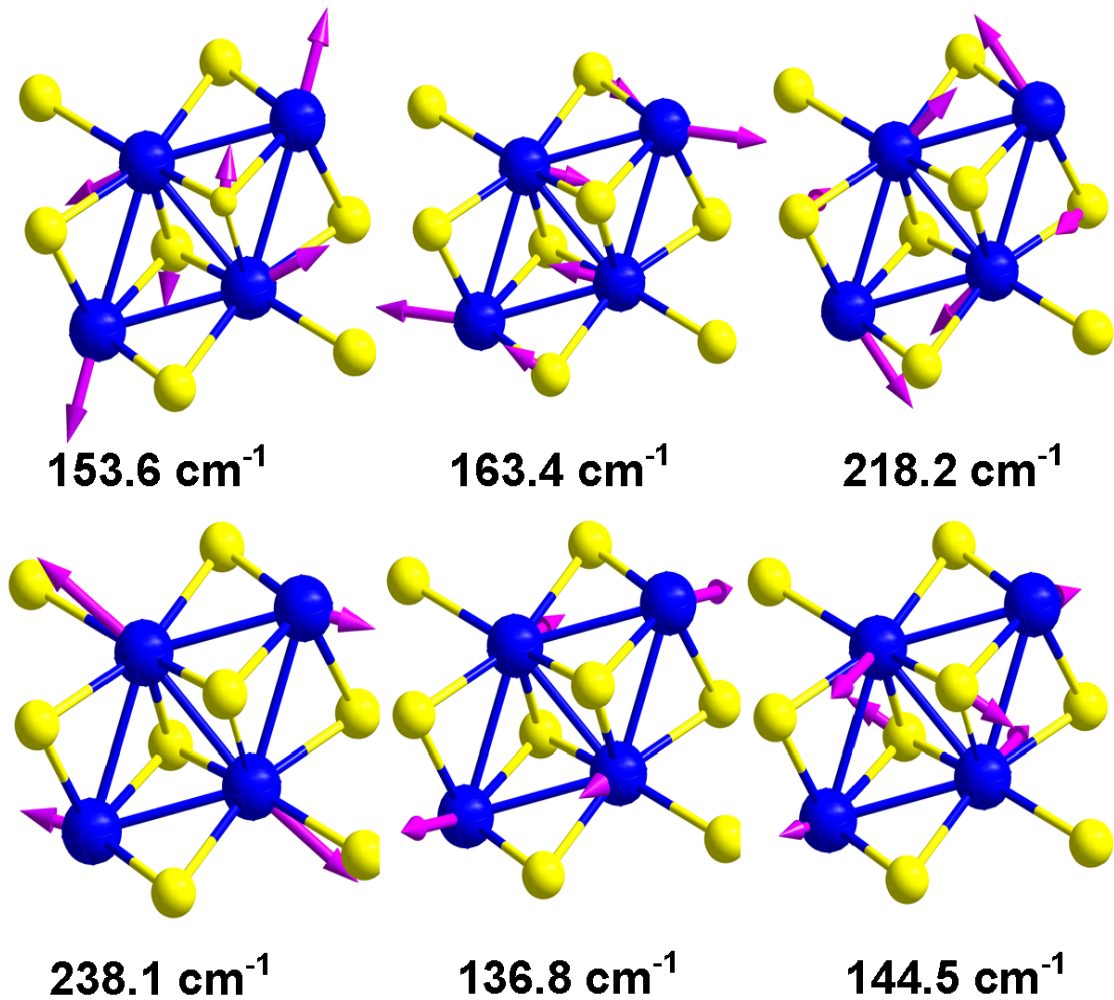

**Table 1.** The relaxed lattice parameters and independent fractional coordinates of bulk ReS$_2$ with DFT-LDA, vdW-DFT and DFT-GGA methods. The experimental data[20] are listed for comparison. The lengths are in units of Å.

| Lattice parameter | a | | b | c | α | β | γ |
|---|---|---|---|---|---|---|---|
| LDA | 6.315 | | 6.482 | 6.415 | 121.4° | 88.3° | 106.6° |
| vdW-DFT | 6.308 | | 6.482 | 6.414 | 121.4° | 88.3° | 106.6° |
| GGA | 7.389 | | 6.588 | 6.527 | 121.3° | 88.1° | 105.0° |
| Exp.[20] | 6.417 | | 6.510 | 6.461 | 121.1° | 88.4° | 106.5° |
| Fractional coordinates | bulk | Re1 | Re2 | S1 | S2 | S3 | S4 |
| x | LDA | 0.503 | 0.492 | 0.210 | 0.276 | 0.696 | 0.761 |
| | vdW-DFT | 0.503 | 0.492 | 0.209 | 0.276 | 0.697 | 0.761 |
| | GGA | 0.502 | 0.492 | 0.252 | 0.309 | 0.667 | 0.722 |
| | Exp.[20] | 0.503 | 0.493 | 0.217 | 0.277 | 0.698 | 0.756 |
| y | LDA | 0.513 | 0.058 | 0.250 | 0.774 | 0.754 | 0.279 |
| | vdW-DFT | 0.513 | 0.058 | 0.250 | 0.774 | 0.755 | 0.279 |
| | GGA | 0.512 | 0.058 | 0.262 | 0.783 | 0.746 | 0.266 |
| | Exp.[20] | 0.511 | 0.056 | 0.250 | 0.771 | 0.753 | 0.273 |
| z | LDA | 0.298 | 0.247 | 0.366 | 0.383 | 0.119 | 0.119 |
| | vdW-DFT | 0.298 | 0.247 | 0.366 | 0.383 | 0.119 | 0.119 |
| | GGA | 0.298 | 0.247 | 0.374 | 0.390 | 0.114 | 0.115 |
| | Exp.[20] | 0.297 | 0.248 | 0.368 | 0.384 | 0.117 | 0.118 |

**Table 2.** The relaxed independent Re-Re bond lengths (the lengths of Re2-Re3 and Re3-Re4 equate to Re1-Re4 and Re1-Re2, respectively a result of inversion symmetry) given by DFT-LDA, vdW-DFT and DFT-GGA methods. The experimental data[20] are listed for comparison. The lengths are in units of Å.

| Re-Re distance | Re1-Re2 | Re1-Re3 | Re1-Re4 |
| --- | --- | --- | --- |
| LDA | 2.77 | 2.68 | 2.78 |
| vdW-DFT | 2.77 | 2.68 | 2.78 |
| GGA | 2.82 | 2.72 | 2.84 |
| Exp.[20] | 2.79 | 2.69 | 2.82 |

**Table 3.** The 18 Raman active frequencies (in units of cm$^{-1}$) in bulk ReS$_2$ of experimental measurement under 633 nm solid state excitation wavelength laser and first-principles theoretical calculations. The A$_g$-like, E$_g$-like and Cp modes are also marked correspondingly.

| Symmetry | Bulk Raman frequency | | | | 1L Raman frequency | | |
|---|---|---|---|---|---|---|---|
| | Exp. | LDA | vdW-DFT | GGA | Exp. | LDA | GGA |
| A$_g$-like | 140.3 | 136.8 | 137.2 | 129.4 | 139.2 | 132.7 | 129.3 |
| A$_g$-like | 145.9 | 144.5 | 145.3 | 137.1 | 145.3 | 142.4 | 137.2 |
| E$_g$-like | 153.1 | 153.6 | 154.0 | 148.0 | 153.6 | 155.5 | 148.3 |
| E$_g$-like | 163.6 | 163.4 | 163.6 | 158.4 | 163.6 | 164.3 | 158.4 |
| E$_g$-like | 217.2 | 218.2 | 218.3 | 208.7 | 217.7 | 220.2 | 208.9 |
| E$_g$-like | 237.1 | 238.1 | 238.0 | 228.4 | 237.7 | 241.3 | 228.7 |
| Cp | 278.3 | 275.1 | 276.4 | 261.5 | 278.3 | 275.6 | 261.9 |
| Cp | 284.2 | 282.6 | 283.5 | 268.5 | 284.7 | 282.8 | 268.8 |
| E$_g$-like | 307.8 | 308.5 | 311.9 | 295.8 | 307.8 | 309.6 | 295.9 |
| E$_g$-like | 311.0 | 312.1 | 312.9 | 298.3 | 311.0 | 311.8 | 298.2 |
| Cp | 320.6 | 318.3 | 319.8 | 303.2 | 320.6 | 317.4 | 303.5 |
| Cp | 324.9 | 325.8 | 327.2 | 311.1 | 324.9 | 326.9 | 311.1 |
| Cp | 348.8 | 349.7 | 351.3 | 332.4 | 348.8 | 350.4 | 332.7 |
| Cp | 368.9 | 370.4 | 372.8 | 354.5 | 369.5 | 371.6 | 354.7 |
| Cp | 377.9 | 381.3 | 382.8 | 363.0 | 377.4 | 380.3 | 363.3 |
| Cp | 407.3 | 408.7 | 412.0 | 393.4 | 408.3 | 410.8 | 393.5 |
| $A_g$ like | 418.7 | 422.3 | 425.1 | 406.7 | 419.3 | 423.8 | 406.8 |
| $A_g$ like | 438.0 | 443.4 | 446.1 | 425.1 | 437.5 | 443.7 | 424.7 |

**Table 4.** The relaxed lattice parameters of monolayer $ReS_2$ with DFT-LDA and DFT-GGA methods. The VASP GGA-PBE results by Tongay et al.[17] and Horzum et al.[42] are listed for comparison. The lengths are in units of Å.

| Lattice parameter | a | b | c | α | β | γ |
|---|---|---|---|---|---|---|
| LDA | - | 6.477 | 6.408 | 121.4° | 88.1° | 106.2° |
| GGA | - | 6.587 | 6.526 | 121.3° | 88.1° | 107.0° |
| GGA[17] | - | 6.51 | 6.41 | - | - | - |
| GGA[42] | - | 6.51 | 6.40 | - | - | - |

ACKNOWLEDGMENT

  The work is supported by the National Key Project for Basic Research of China (Grants No. 2011CB922101, 2015CB921600, 2013CBA01603), the National Natural Science Foundation of China (Grants No. 91122035, 11374142, 11174124, 11374137, 11474150), the Natural Science Foundation of Jiangsu Province (BK20130544, BK20140017), the Specialized Research Fund for the Doctoral Program of Higher Education (20130091120040), and Fundamental Research Funds for the Central Universities. H.T.Y. was supported by the Department of Energy, Office of Basic Energy Sciences, Division of Materials Sciences and Engineering, under contract DE-AC02-76SF00515. The project is also funded by Priority Academic Program Development of Jiangsu Higher Education Institutions. We also acknowledge the support for the computational resources by the High Performance Computing Center of Nanjing University.